\documentclass{elsart}

\usepackage{amssymb}
\usepackage{graphicx}
\usepackage{latexsym}
\usepackage{epstopdf}
\usepackage{epsfig}

\begin{document}

\begin{frontmatter}

\title{Comparison of hyperfine anomalies in the $5S_{1/2}$ and $6S_{1/2}$  levels
of $^{85}$Rb and $^{87}$Rb.}

\author[label1]{A. P\'{e}rez Galv\'{a}n}
\author[label1]{Y. Zhao}
\author[label1]{L. A. Orozco\corauthref{cor1}}
\ead{lorozco@umd.edu}
\author[label2]{E. G\'{o}mez \thanksref{label3}}
\author[label2]{A. D. Lange \thanksref{label4}}
\author[label2]{F. Baumer \thanksref{label5}}
\author[label2]{G. D. Sprouse}
\corauth[cor1]{Corresponding Author}
\address[label1]{Joint Quantum Institute, Department of Physics,
University of Maryland and National Institute of Standards and Technology, College Park, MD 20742-4100, USA.}
\address[label2]{Department of Physics and Astronomy, SUNYSB, Stony Brook NY 11794-3800, USA.}
\thanks[label3]{Present address: Instituto de F\'{i}sica,
Universidad Aut\'{o}noma de San Luis Potos\'{i}, SLP, M\'{e}xico.}
\thanks[label4]{Present address: Institut f\"{u}r
Experimentalphysik, Universit\"{a}t Innsbruck, Austria.}
\thanks[label5]{Present address: Institut f\"{u}r
Experimentalphysik, Heinrich-Heine-Universit\"{a}t D\"{u}sseldorf,
Germany.}
\begin{abstract}
We observe a hyperfine anomaly in the measurement of the hyperfine
splitting of the $6S_{1/2}$ excited level in rubidium. We perform
two step spectroscopy using the
$5S_{1/2}\rightarrow5P_{1/2}\rightarrow6S_{1/2}$ excitation
sequence. We measure the splitting of the $6S_{1/2}$ level and
obtain for the magnetic dipole constants of $^{85}$Rb and
$^{87}$Rb  $A=239.18(4)$~MHz and $A=807.66(8)$~MHz, respectively.
The hyperfine anomaly difference of $_{87}\delta_{85}=-0.0036(2)$
comes from the Bohr Weisskopf effect: a correction to the point
interaction between the finite nuclear magnetization and the
electrons, and agrees with that obtained in the $5S_{1/2}$ ground
state.
\end{abstract}

\end{frontmatter}

Measurements of hyperfine splittings provide a low energy approach
to study the magnetic moment distribution in the nucleus. The
hyperfine splitting is due mainly to the coupling between the
magnetization of the nucleus with the magnetic field created by
the electrons. To a very good approximation the coupling is taken
to be point-like. However, high precision measurements of the
splitting can reveal deviations from the point interaction. These
deviations, also called hyperfine anomalies, come from considering
how finite magnetic and charge distributions affect the
interaction. Bohr and Weisskopf \cite{bohr50} first discussed the
finite magnetization effect in the anomaly. The modified charge
potential that the valence electron sees as it gets closer to the
nucleus, the Breit-Crawford-Rosenthal-Schawlow
\cite{rosenthal32,crawford49} effect, is the other source for a
hyperfine anomaly.

We show in this letter a measurement of the excited state
hyperfine splittings in two isotopes precise enough to extract a
difference in the hyperfine anomalies. The hyperfine anomaly
differences of the ground state in francium were measured by our
group \cite{grossman99} by comparing the ratio of the hyperfine
splittings of the $7P_{1/2}$ (which has a small anomaly) and the
$7S_{1/2}$ (which has the large anomaly) \cite{persson98}. These
measurements represent excellent benchmarks to test
state-of-the-art \emph{ab initio} calculations of the electronic
and nuclear wave functions for future parity non-conservation
measurements \cite{gomez06,gomez07}. They give the possibility to
look at how the systematic addition of a neutron modifies the
electronic wavefunction in the nucleus, and how the nucleus
rearranges itself. As a new generation of proposed parity
violation experiments in atoms starts,
\cite{gomez06,bouchiat97,bouchiat07}, it is important to
understand the nuclear structure limiting factors for the
extraction of weak coupling parameters from the measurements
\cite{derevianko02}. This letter presents a new experimental
approach to extract and study such effects in chains of isotopes.

The hyperfine shift for these levels is given by
$E_{HF}={A}(F(F+1)-I(I+1)-J(J+1))/2$ where $A$ is the magnetic
dipole constant, $F$ is the total angular momentum, $I$ is the
nuclear spin, and $J$ is the total electronic angular momentum.
Derivations of $A$ assume that the nucleus is a point particle
with magnetic moment $\mu_{N}$ and charge $Z$. However, this is
not the case. The nucleus has a finite extension and structure.
The value of the magnetic dipole constant has to be modified to
include the effect of the charge and magnetic distribution on the
electronic wave function. We can write, following
Ref.~\cite{kopfermann} $A$ for an extended ($ext$) nucleus as a
function of the point value ($point$) (Eq. \ref{Aext}):
\begin{eqnarray}
A_{point}&=&\frac{16\pi}{3} \frac{\mu_{0}}{4\pi h}g_{I}\mu_{N}
\mu_{B}|\psi(0)|^{2} \nonumber f_{R}, \nonumber \\
A_{ext}&=&A_{point}(1+\epsilon_{BCRS})(1+\epsilon_{BW}),
\label{Aext}
\end{eqnarray}
where $\psi(0)$ is the electronic wave function evaluated at the
center of the nucleus, $\mu_{B}$ is the Bohr magneton, $\mu_{N}$
is the nuclear magneton, $g_{I}$ is the nuclear g-factor, and
$f_{R}$ represents the relativistic enhancement. The last two
terms of this expression modify the hyperfine interaction to
account for an extended nucleus. The first of them
($\epsilon_{BCRS}$), the Breit-Crawford-Rosenthal-Schawlow (BCRS)
correction, accounts for the fact that the valence electron sees a
modified potential as it gets closer to the nucleus. The second
one ($\epsilon_{BW}$), the Bohr Weiskopf (BW) effect, focuses on
the magnetic distribution of the nucleus.

The fractional difference in the mean charge radius between the
two isotopes of interest ($^{85}$Rb and $^{87}$Rb) is less than
$10^{-3}$ and justifies neglecting the BCRS correction
\cite{thibault81,angeli04}. $^{87}$Rb has a closed-neutron shell
that makes the nuclear charge distribution insensitive to the
addition or subtraction of neutrons \cite{thibault81}. A
calculation of $\epsilon_{BCRS}$ \cite{armstrong} gives a value of
1\% compared to the point nucleus; however, since we are
interested in comparing the two anomalies we find, using the
values of the charge radius of Ref.~\cite{angeli04}, the
difference between the corrections of both isotopes is less than
$10^{-4}$.

The BW correction describes the modification of the hyperfine
interaction due to a finite distribution of nuclear magnetization.
The addition of more neutrons to the nucleus yields a different
spatial magnetization. The magnitude of this correction for an
individual nucleus is smaller than the BCRS correction but depends
heavily on the spin and orbital angular momentum of the nucleus
and not just on its radius. For the particular case of $^{85}$Rb
and $^{87}$Rb the addition of two neutrons changes the value of
the valence proton from an $f_{5/2}$ orbital to a $p_{3/2}$
orbital. The lighter of the isotopes, $^{85}$Rb, has the spin
angular momentum of the valence proton anti-aligned with its
orbital angular momentum while $^{87}$Rb adds them both together.
The two neutron holes deform the nucleus very slightly and change
the order of the orbitals that are almost degenerate. The
difference of the hyperfine anomaly contributions
$_{87}\delta_{85}=\epsilon^{87}_{BW}-\epsilon^{85}_{BW}$ is
dominated by the BW contribution as we take the BCRS correction as
equal for both isotopes.

Since the magnetic moments of the nuclei are well known
\cite{duong93} we can use a ratio to extract the Bohr-Weisskopf
effect difference:
\begin{equation}
\frac{A^{87}g^{85}}{A^{85}g^{87}}\simeq 1+_{87}\delta_{85}.
\label{anom}
\end{equation}

\begin{figure}
\leavevmode \centering
  \includegraphics[width=3in]{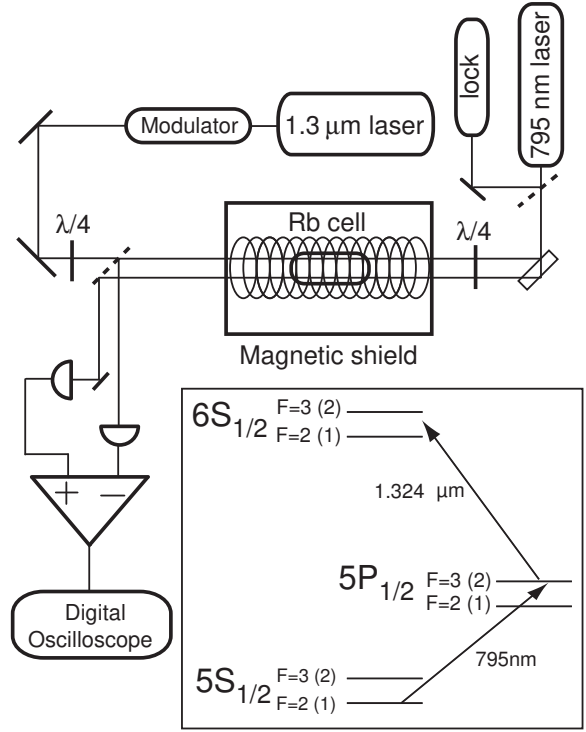}
  \caption{Schematic of the experimental apparatus. The inset shows the
  energy levels relevant to our experiment (not drawn to scale)
  with quantum numbers corresponding to $^{85}$Rb ($^{87}$Rb).}
  \label{figure1}
\end{figure}

Because the BW corrections are small, we require a method to
measure the hyperfine splitting with good precision. We perform
two-photon absorption spectroscopy in a 30 cm long 2.5 cm diameter
Rb cell with natural isotopic abundances to measure the hyperfine
splitting of the $6S_{1/2}$ level using the $5P_{1/2}$ as an
intermediate level (see Fig.~\ref{figure1} for the block diagram
of the experimental setup and the atomic energy levels involved).
The cell has no buffer gas and is in a controlled magnetic
environment.

We monitor the change in absorption of a laser resonant with the
transition to the $5P_{1/2}$ level at 795 nm as we scan the 1.3
$\mu$m laser over the hyperfine structure of the $6S_{1/2}$ level.
We modulate the 1.3 $\mu$m laser at a frequency close to half of
the hyperfine separation to produce sidebands well separated in
energy. This way we only need to scan the laser a small range
around the midpoint between the two hyperfine levels of the
$6S_{1/2}$ level such that the sidebands scan over both hyperfine
levels. The sideband signals appear in the absorption profile and
work as an \emph{in situ} scale. We record absorption profiles at
different values for the modulation to find, with the help of
linear regression plots, the point where the sidebands cross the
midpoint of the two main peaks. Typical fits give correlation
coefficients that differ from unity by at most $2 \times 10^{-4}$.

A Titanium Sapphire  laser with a linewidth better than 500 kHz
tuned to the  Rb $D_1$ line at 795 nm excites the first step of
the transition. A Pound-Drever-Hall setup on an independent Rb
cell locks the 795 nm laser to the F=1(2)$\rightarrow$F'=2(3)
transition in $^{87}$Rb $(^{85}$Rb). A grating narrowed diode
laser at 1.324 $\mu$m with a linewidth better than 500 kHz excites
the second transition. We scan the frequency of the 1.3~ $\mu$m
laser using a triangular shaped voltage ramp from a synthesized
function generator at 4 Hz applied to the piezo control of the
grating and monitor its frequency with a wavemeter. A
fiber-coupled semiconductor optical amplifier increases the power
of the 1.3 $\mu$m laser before it goes to a large bandwidth
($\approx 10$ GHz) Electro-Optic Modulator (EOM). A signal
generator modulates this EOM in the range between 100 and 900 MHz
to imprint the sidebands necessary for the measurement.

Before going through the rubidium vapor cell the laser beams are
circularly polarized by $\lambda/4$ waveplates. A counter
propagating 1.3 $\mu$m laser beam with a power of up to 4 mW
overlaps one of the 795 nm beams. The relative power on the
sidebands is less than a tenth of the total and its waist is 1 mm.
The lasers overlap to a precision of better than 1 mm along 75 cm.
The cell resides in the center of a 500-turn solenoid (7.4
Gauss/A) contained inside a three layered magnetic shield. The
solenoid is 70 cm long and has a diameter of 11.5 cm.

A thick glass plate splits the 795 nm laser beam into two
co-propagating beams. The power of each beam is approximately
10~$\mu$W with a waist of 1 mm. We find that doing the two step
excitation in either a ($\sigma^{+},\sigma^{-}$) or
($\sigma^{-},\sigma^{+}$) polarization sequence for the 795 nm and
1.3 $\mu$m lasers respectively, increases the probability of the
transition to the $6S_{1/2}$ level, and allows us to check for
optical pumping and Zeeman shift effects. We place the Rb cell in
a uniform magnetic field convenient for the $\Delta m_{F}=\pm1$
transitions. The addition of the magnetic field has the two-fold
advantage of providing a quantization axis and a tool to probe
systematic effects. We measure the hyperfine splitting for
different values, including reversals, of the magnetic field and
polarization. We can then extract by interpolation the value of
the hyperfine splitting at zero magnetic field.

\begin{figure}[h]
\leavevmode \centering
   \includegraphics[width=3in]{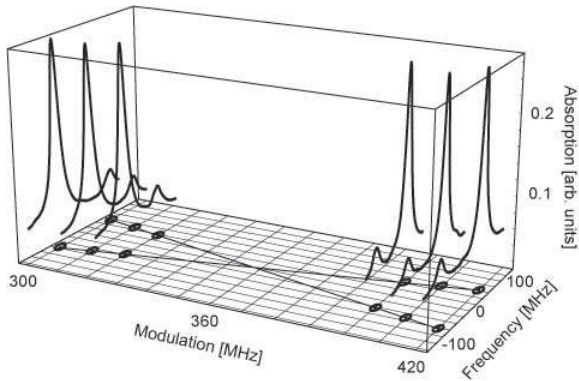}
  \caption{Experimental traces that illustrate sideband
  crossing for $^{85}$Rb. The intersection of the two projected lines
marks half the hyperfine separation.}
  \label{figure2}
\end{figure}

After the glass cell an independent photodiode detects each 795 nm
laser beam. The output of both detectors goes to a differential
amplifier to reduce common noise. A digital storage oscilloscope
records the signal and averages for about three minutes. The
stability of the 1.3 $\mu$m laser center frequency over this time
is better than 0.3 MHz as measured by a Fabry Perot cavity. The
optical attenuation at line center ($D_1$ line) is 0.4 for
$^{85}$Rb. Fig. \ref{figure2} presents an example of the
oscilloscope traces for different sideband frequencies. The
projection shows the positions of the peaks joined by the linear
interpolation, the error in the position is much smaller than the
dots. We obtain signal to noise ratios in excess of 100 for the
sideband resonances.

We analyze how the following parameters influence the hyperfine
separation: the peak shape model for the non-linear fit to obtain
the separation of the centers of the lines, scan width and
repetition rate of the 1.3 $\mu$m laser, power of the 795 nm and
1.3 $\mu$m lasers. We look for optical pumping effects, magnetic
field effects, and temperature, though the vapor density, that can
influence the hyperfine separation. We have only been able to put
upper bounds on the systematic shifts within our resolution and we
determine that statistical fluctuations, as stated by the standard
error of the mean, dominate the uncertainty of the hyperfine
splitting ($\nu_{HF}$) measurement as seen in Table \ref{Table 1}.

\begin{table}[b]
  \leavevmode \centering
   \begin{tabular}{lcc}
   Systematic effects       &$\nu_{HF}$($^{85}$Rb) (MHz)&$\nu_{HF}$($^{87}$Rb) (MHz)\\ \hline
   Optical pumping effects &        $\leq$ 0.016   & $\leq$ 0.029 \\
   Power of 795 nm laser   &        $\leq$ 0.020   & $\leq$ 0.005  \\
   Power of 1.3 $\mu$m laser  &     $\leq$ 0.011   & $\leq$ 0.011  \\
   Atomic density          &        $\leq$ 0.020   & $\leq$ 0.010  \\
   Non linear fit          &        $\leq$ 0.028   & $\leq$ 0.023  \\
   B-field fluctuations    &        $\leq$ 0.015   & $\leq$ 0.025  \\\hline
   Total Systematic        &        $\leq$ 0.047   & $\leq$ 0.047  \\\hline
   Statistical error       &        0.100          & 0.160  \\
   \textbf{TOTAL}           &        \textbf{0.110 }         & \textbf{0.167}  \\

   \end{tabular}
  \caption{Error budget for the hyperfine splitting measurement}
  \label{Table 1}
\end{table}

Table \ref{Table 2} presents the results of the hyperfine
splitting of the $6S_{1/2}$ level  as well as the corresponding
values of the magnetic dipole constants for both isotopes. Our
results are consistent with the past measurements of Gupta {\it et
al.} \cite{gupta73} and represent a precision improvement by a
factor of 63 for $^{87}$Rb and by a factor of 30 for $^{85}$Rb.
The table contains the theoretical predictions for $^{85}$Rb  of
Safronova {\it et al.} \cite{safronova99}. The theoretical
calculation includes corrections for the finite size of the
nuclear magnetic moment distribution, which is modelled as a
uniformly magnetized ball. The theory value for $^{87}$Rb comes
from multiplying the theoretical value of the splitting in
$^{85}$Rb by the ratio of nuclear gyromagnetic ratios and does not
include a hyperfine anomaly difference.

\begin{table}[h]
  \leavevmode \centering
   \begin{tabular}{lcc}
                 & $^{85}$Rb [MHz] & $^{87}$Rb [MHz] \\\hline
   $\nu_{HF}$ this experiment   &   717.540(110)    & 1615.320(167) \\
   $A$ previous experiment \cite{gupta73}&   239.3(12)    & 809.1(50) \\
   $A$ this experiment          &   239.18(4)    & 807.66(8) \\
   $A$ theory \cite{safronova99} &   238.2   & 807.3    \\\hline
   \end{tabular}
  \caption{Hyperfine splittings and magnetic dipole constants for the $6S_{1/2}$ level.}\
  \label{Table 2}
\end{table}

Optical pumping effects are the most delicate of all the
systematic effects. The polarization of the lasers determine the
relative sizes of the peaks ($m$ sublevels) that form  the
resonances of the $6S_{1/2}$ hyperfine levels. For our
experimental conditions (around 1 G with circularly polarized
light) we observe no difference between positive and negative
magnetic field directions. Table \ref{Table 1} shows the bound for
this effect.

We change the power of the 795 nm  and 1.3~$\mu$m lasers to look
for any kind of dependence that would indicate line splitting
effects such as the Autler-Townes effect or power broadening.  We
observe no systematic variation in the hyperfine separation as a
function of laser powers. No systematic effect is found in the
hyperfine splitting for changes in the temperature of the cell
from 23 to 40 $^{\circ}$C. The bounds for these effects are on
Table \ref{Table 1}.

The observed linewidths range from 30 to 40 MHz and fit (reduced
$\chi^{2}$) Lorentzian profiles better ($\approx 1$) than Gaussian
($\approx 10$), with the Voigt profile between the two. We do not
observe changes in the splitting that depend on the frequency
range fitted around the resonances. The values obtained for the
difference between the centers of the resonances using Gaussian
fits give consistent results with the Lorenzian fits, not changing
the measured splitting by more than the reported error.

The three layers of magnetic shielding limit the external magnetic
fluctuations to less than 1 mG, while the stability of the
magnetic field from the solenoid is comparable. We operate within
the low magnetic field regime (1 G) and extrapolate to B=0 for
both polarization sequences with linear fits with an average
correlation coefficient of 0.97.

\begin{figure}[h]
\leavevmode \centering
   \includegraphics[width=3in]{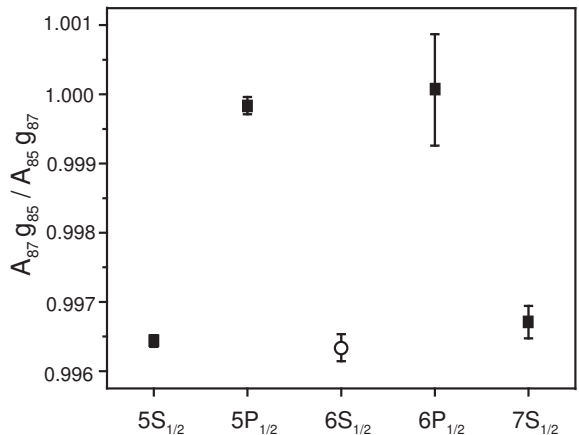}
  \caption{Ratio of hyperfine constants normalized by the nuclear g factors showing hyperfine anomaly differences in $^{85}$Rb
  and $^{87}$Rb based on five different electronic
  states. The value for the $6S_{1/2}$ (circle) comes from the present
  measurement. See the text for the other references (squares).}
  \label{figure3}
\end{figure}

Figure \ref{figure3} shows the normalized ratio of hyperfine
constants from Eq.~\ref{anom} to show the hyperfine anomaly
difference extracted from our measurement in the $6S_{1/2}$ level
of $^{85}$Rb and $^{87}$Rb as well as the corresponding values for
other levels currently in the literature. We use the values from
Duong {\it et al.} \cite{duong93} to obtain the ratio of $g$
values $g_{85}/g_{87}=0.295061 \pm 0.00003$, which is consistent
with the values quoted by Stone \cite{stone05}, to calculate the
hyperfine anomaly difference from the measured $A$ coefficients.
The hyperfine splittings of the $5S_{1/2}$ ground state are very
well known as they are used in atomic clocks \cite{arimondo77}.
The $5P_{1/2}$ ratio comes from the work of Barwood {\it et al.}
\cite{barwood91} which is consistent with more recent frequency
comb measurements \cite{marian04,marian05}. The $6S_{1/2}$ ratio
corresponds to the present measurement which gives
$_{87}\delta_{85}(6S_{1/2})=-0.0036(2)$. The difference in the
anomalies is indeed a factor of thirty larger than the expected
BCRS contribution and it comes from the BW effect. The $6P_{1/2}$
ratio comes from the critical evaluation of Arimondo {\it et al.}
\cite{arimondo77}, while we find from the recent frequency comb
measurement of the $7S_{1/2}$ state by Chui {\it et al.} \cite
{chui05} $_{87}\delta_{85}(7S_{1/2})=-0.0033(2)$, although they do
not calculate nor mention the hyperfine anomaly in their paper. We
have not been able to find in the literature values for higher
levels with adequate precision to include them in the figure. Fig.
\ref{figure3} shows that the hyperfine anomaly difference measured
with the $S_{1/2}$ states is independent of the principal quantum
number, just as Bohr and Weisskopf predicted \cite{bohr50}. The
measurements are consistent with each other as are the $P_{1/2}$
states that show a much smaller anomaly.


We have measured the hyperfine splittings of the $6S_{1/2}$ level
of $^{85}$Rb and $^{87}$Rb to a precision of 153 ppm and 103 ppm,
respectively.  Our measurement allows us to extract the hyperfine
anomaly difference of the two isotopes with a precision of 5\%. It
is in excellent agreement with that of the ground state,
confirming that it is the angular momentum and not the principal
quantum number that enters in the Bohr Weisskopf effect
\cite{bohr50}. We show also that nuclear magnetization information
can be extracted through careful probing of excited state
hyperfine splittings.

Work supported by NSF.  E. G. acknowledges support from CONACYT.

%
%





%


\begin{thebibliography}{10}
\expandafter\ifx\csname url\endcsname\relax
  \def\url#1{\texttt{#1}}\fi
\expandafter\ifx\csname
urlprefix\endcsname\relax\def\urlprefix{URL }\fi

\bibitem{bohr50}
A.~Bohr, V.~F. Weisskopf,
  Phys. Rev. 77 (1950) 94.

\bibitem{rosenthal32}
J. E. Rosenthal, G. Breit,
  Phys. Rev. 41 (1932) 459.

\bibitem{crawford49}
M.~F. Crawford, A.~L. Schawlow,
  Phys. Rev. 76 (1949) 1310.

\bibitem{grossman99}
J.~S. Grossman, L.~A. Orozco, M.~R. Pearson, J.~E. Simsarian,
G.~D. Sprouse,
  W.~Z. Zhao,
  Phys. Rev. Lett. 83 (1999) 935.

\bibitem{persson98}
J.~R. Persson, Eur. Phys. J. A 2 (1998) 3.

\bibitem{gomez06}
E.~Gomez, L.~A. Orozco, G.~D. Sprouse,
  Rep. Prog. Phys. 66
  (2006) 79.

\bibitem{gomez07}
E.~Gomez, S.~Aubin, L.~A. Orozco, G.~D. Sprouse, D.~DeMille,
  Phys. Rev. A 75
  (2007) 033418.

\bibitem{bouchiat97}
M.-A. Bouchiat, C.~Bouchiat,
Rep. Prog.
Phys. 60
  (1997) 1351.

\bibitem{bouchiat07}
M.~A. Bouchiat,
  Phys.
  Rev. Lett. 98 (2007) 043003.

\bibitem{derevianko02}
A.~Derevianko, S.~G. Porseu,
  Phys. Rev. A
  65 (2002) 052115.

\bibitem{kopfermann}
H.~Kopfermann, Nuclear Moments, Academic Press, New York, 1958.

\bibitem{thibault81}
C.~Thibault, F.~Touchard, S.~B{\"u}ttgenbach, R.~Klapisch,
M.~{de~Saint~Simon},
  H.~T. Duong, P.~Jacquinot, P.~Juncar, S.~Liberman, P.~Pillet, J.~Pinard,
  J.~L. Vialle, A.~Pesnelle, G.~Huber,
  Phys. Rev. C 23
  (1981) 2720.

\bibitem{angeli04}
I.~Angeli,
  At. Data Nucl. Data Tables 87 (2004) 185.

\bibitem{armstrong}
L.~{Armstrong~Jr}, Theory of the Hyperfine Structure of Free
Atoms, Wiley-Interscience, New York, 1971.

\bibitem{duong93}
H.~T. Duong, C.~Ekstr{\"{o}}m, M.~Gustafsson, T.~T. Inamura,
P.~Juncar,
  P.~Lievens, I.~Lindgren, S.~Matsuki, T.~Murayama, R.~Neugart, T.~Nilsson,
  T.~Nomura, M.~Pellarin, S.~Penselin, J.~Persson, J.~Pinard, I.~Ragnarsson,
  O.~Redi, H.~H. Stroke, J.~L. Vialle, the {ISOLDE}~Collaboration, Nuc. Instr.
  and Meth. A 325 (1993) 465.

\bibitem{safronova99}
M.~S. Safronova, W.~R. Johnson, A.~Derevianko,
  Phys. Rev. A 60
  (1999) 4476.

\bibitem{gupta73}
R.~Gupta, W.~Happer, L.~K. Lam, S.~Svanberg,
  Phys. Rev. A 8 (1973) 2792.

\bibitem{stone05}
N.~J. Stone,
  At. Data Nucl. Data Tables 90 (2005) 75.

\bibitem{arimondo77}
E.~Arimondo, M.~Inguscio, P.~Violino,
  Rev. Mod. Phys. 49 (1977) 31.

\bibitem{barwood91}
G.~P. Barwood, P.~Gill, W.~R.~C. Rowley,
  Appl. Phys. B:
  Photophys. Laser Chem. 53 (1991) 142.

\bibitem{marian04}
A.~Marian, M.~C. Stowe, J.~R. Lawall, D.~Felinto, J.~Ye,
  Science 306 (2004) 2063.

\bibitem{marian05}
A.~Marian, M.~C. Stowe, D.~Felinto, J.~Ye,
  Phys. Rev.
  Lett. 95 (2005) 023001.

\bibitem{chui05}
H.-C. Chui, M.-S. Ko, Y.-W. Liu, J.-T. Shy, J.-L. Peng, H.~Ahn,
  Opt. Lett. 30 (2005) 842.

\end{thebibliography}
\end{document}